\documentclass[12pt,preprint]{aastex}

\bibpunct[, ]{(}{)}{;}{a}{}{,}
\newcommand{\swift}{{\it Swift}}
\newcommand{\xrt}{{\rm XRT}}
\newcommand{\grb}{GRB~050509B}
\slugcomment{To appear in ApJL}

\shorttitle{Constraints on short GRB models}
\shortauthors{Hjorth et al.}

\begin{document}

\title{
GRB~050509B: Constraints on short gamma-ray burst models% 
\thanks{Based on observations collected at the European
Southern Observatory, Paranal, Chile (ESO Programme 075.D-0261).}
}


\author{ 
J.~Hjorth\altaffilmark{2}, 
J.~Sollerman\altaffilmark{2,3},
J.~Gorosabel\altaffilmark{4}, 
J.~Granot\altaffilmark{5},
S.~Klose\altaffilmark{6},
C.~Kouveliotou\altaffilmark{7}, 
J.~Melinder\altaffilmark{3}, 
E.~Ramirez-Ruiz\altaffilmark{8,9},
R.~Starling\altaffilmark{10}, 
B.~Thomsen\altaffilmark{11},
M.~I.~Andersen\altaffilmark{12},
J.~P.~U.~Fynbo\altaffilmark{2}, 
B.~L.~Jensen\altaffilmark{2},
P.~M.~Vreeswijk\altaffilmark{13}, 
J.~M.~Castro Cer\'on\altaffilmark{2},
P.~Jakobsson\altaffilmark{2}, 
A.~Levan\altaffilmark{14},
K.~Pedersen\altaffilmark{2},
J.~E.~Rhoads\altaffilmark{15},
N.~R.~Tanvir\altaffilmark{16},
D.~Watson\altaffilmark{2},
R.~A.~M.~J.~Wijers\altaffilmark{10}
}


\altaffiltext{2}{Dark Cosmology Centre, Niels Bohr Institute, University of
Copenhagen, Juliane Maries Vej 30, DK-2100 Copenhagen \O, Denmark; 
jens,jfynbo,brian$\_$j,pallja,kp,darach@astro.ku.dk; josemari@alumni.nd.edu.} 
\altaffiltext{3}{Department of Astronomy, Stockholm University, AlbaNova, 
S-106 91 Stockholm, Sweden; jesper,jens@astro.su.se.} 
\altaffiltext{4}{Instituto de Astrof\'\i sica de Andaluc\'\i a 
(IAA-CSIC), P.O. Box 03004, E-18080 Granada, Spain; jgu@iaa.es.}
\altaffiltext{5}{Kavli Institute for Particle Astrophysics and Cosmology, 
Stanford University, P.O. Box 20450, MS 29, Stanford, CA 94309;
granot@slac.stanford.edu.}
\altaffiltext{6}{Th\"uringer Landessternwarte Tautenburg, Sternwarte 5, 
D-07778 Tautenburg, Germany; klose@tls-tautenburg.de.}
\altaffiltext{7}{NASA/Marshall Space Flight Center, National Space Science 
Technology Center, XD-12, 320 Sparkman Drive, Huntsville, AL 35805;
Chryssa.Kouveliotou-1@nasa.gov.}
\altaffiltext{8}{Institute for Advanced Study, Einstein Drive, 
Princeton, NJ 08540; enrico@ias.edu.}
\altaffiltext{9}{Chandra Fellow.}
\altaffiltext{10}{Astronomical Institute `Anton Pannekoek', University of 
Amsterdam, NL-1098 SJ Amsterdam, The Netherlands; 
starling,rwijers@science.uva.nl.}
\altaffiltext{11}{Department of Physics and Astronomy,
University of Aarhus, DK-8000 \AA rhus C, Denmark; bt@phys.au.dk.}
\altaffiltext{12}{Astrophysikalisches Institut Potsdam, An der Sternwarte 16, 
D-14482 Potsdam, Germany; mandersen@aip.de.}
\altaffiltext{13}{European Southern Observatory, Alonso de C\'ordova 
3107, Casilla 19001, Santiago 19, Chile; pvreeswi@eso.org.}
\altaffiltext{14}{Department of Physics and Astronomy, 
University of Leicester, University Road, Leicester LE1 7RH, UK;
anl@star.le.ac.uk.}
\altaffiltext{15}{Space Telescope Science Institute, 
3700 San Martin Drive, Baltimore, MD 21218; rhoads@stsci.edu.}
\altaffiltext{16}{Centre for Astrophysics Research, 
University of Hertfordshire, College Lane, Hatfield AL10 9AB, UK;
nrt@star.herts.ac.uk.}

\begin{abstract}
  We have obtained deep optical images with the Very Large Telescope
  at ESO of the first well-localized short-duration gamma-ray burst,
  \grb. From $V$ and $R$ imaging, initiated $\sim2$ days after
  the GRB trigger and lasting up to three weeks, we detect no variable
  object inside the small \swift/XRT X-ray error circle down to 
  $2\sigma$ limits of $V=26.5$ and $R=25.1$.
  The X-ray error circle includes a giant elliptical galaxy
  at $z=0.225$, which has been proposed as the likely host of this
  GRB. Our limits indicate that if the GRB originated at $z=0.225$,
  any supernova-like event accompanying the GRB would have to be over
  100 times fainter than normal Type Ia SNe or Type Ic hypernovae, 5
  times fainter than the faintest known Ia or Ic SNe, and fainter than
  the faintest known Type II SNe. Moreover, we use the optical limits 
  to constrain the energetics of the GRB outflow. Simple models 
  indicate that, unless the intrinsic energy in the outflow from \grb\ was 
  $\ll 10^{51}\;$erg, there was very little radioactive material with 
  efficient decay timescales for generating a large luminosity.
  These limits strongly constrain progenitor models for this short GRB.
\end{abstract}


%% subject header.  Objects should be in the appropriate "individual"
%% headers (e.g. quasars: individual, stars: individual, etc.) with the
%% additional provision that the total number of headers, including each
%% individual object, not exceed six.  The \objectname{} macro, and its
%% alias \object{}, is used to mark each object.  The macro takes the object
%% name as its primary argument.  This name will appear in the paper
%% and serve as the link's anchor in the electronic edition if the name
%% is recognized by the data centers.  The macro also takes an optional
%% argument in parentheses in cases where the data center identification
%% differs from what is to be printed in the paper.



\section{Introduction\label{introduction}}

While it is now well established that long-duration $\gamma$-ray
bursts (GRBs) coincide with the explosions of massive stars leading to
very energetic core-collapse supernovae (SNe)
\citep{galama98,macfadyen99,bloom99,stanek03,hjorth03}, the origin of
short/hard GRBs, characterized as having durations $< 2$~s 
and hard spectra \citep{kouveliotou93}, remains unknown. There have been
as yet no afterglow detections in the very few cases where searches
for optical counterparts of short GRBs were performed, primarily due
to the lack of early and precise localizations
\citep{kehoe01,gorosabel02,hurley02,klotz03}.

Recently, the \swift\ satellite \citep{gehrels04} provided the first
rapid and accurate X-ray localization of a short/hard GRB, opening the
window for rapid progress on the origin of short GRBs.  \grb\
\citep{gehrels05} was detected on 2005 May 09 at 04:00:19.23 (UT)
by the \swift\ Burst Alert Telescope
({\rm BAT}). It was a short (40~ms) and fairly hard burst.  The
\swift\ X-ray Telescope (\xrt) determined the source location to be
R.A. = 12$^{\rm h}$ 36$^{\rm m}$ 13\fs58, decl. = $+28^{\circ}$
59$^{\prime}$ 01\farcs3 \citep[J2000, error radius 9\farcs3,][]{gehrels05}.

The error region of \grb\ was observed by several groups \citep[see,
e.g.,][]{bloom05,gehrels05,cenko05,alberto05}.  Remarkably, the burst error
circle overlaps with a giant elliptical galaxy,
2MASX~J$12361286+2858580$ (hereafter G1) at $z = 0.225$ (see
Fig.~\ref{f:allimages} and \citealt{bloom05}), belonging to the
cluster of galaxies NSC~J123610+285901
\citep{gal03}.  The probability of a
chance alignment of a GRB and such a galaxy is
of order $\sim10^{-3}$. Assuming that G1 is therefore 
the host galaxy, \citet{bloom05} and \citet{gehrels05} argued that a
likely origin of \grb\ is a neutron star (NS)-NS or NS-black hole (BH)
merger.

It should be noted that a merger does not necessarily 
imply the absence of optical
or other long-wavelength phenomena after the GRB.  For example, the
`mini-SN' model \citep{LP98} predicts a bright optical flash of
much shorter duration than the one from a `normal' SN,
typically of about one day. But there are alternative scenarios for
the origin of short GRBs. \citet{zhang03} have suggested that short
GRBs may be a variant of long GRBs, e.g., `collapsar'-like events
leading to stripped-core, core-collapse SNe, much like those
seen in conjunction with long GRB afterglows \citep[see
also][]{ghirlanda04,yamazaki04}.  
%We note that two of the three
%spectroscopically confirmed long GRB-SN associations to date (GRB
%980425/SN 1998bw at $z=0.0085$ and GRB 031203/SN 2003lw at $z=0.106$)
%were detected in the optical because of their very strong Type Ic
%SNe rather than their afterglows
%\citep{galama98,prochaska04,malesani04,thomsen04}.  
An alternative
suggestion is that short GRBs are related to thermonuclear explosions,
leading to Type Ia SNe \citep{dar04,dado05}.  Finally,
\citet{germany00} even suggested that the peculiar 
Type II SN~1997cy was
related to the short GRB~970514 based on their temporal
and spatial coincidence. It is obvious from the above, that a search
for a SN associated with \grb\ would help constrain both the
energetics of short GRBs and, possibly, their progenitor models
\citep{fan05}.

We have, therefore, obtained deep images of the \xrt\ error circle at
the expected peak time of the putative SN, as well as early
images for comparison. In this Letter we present our observations and
analysis (\S~\ref{observations}) and discuss the constraints these set
on short GRB energetics and progenitor models (\S~3).  A cosmology
with $H_0=70$\,km\,s$^{-1}$\,Mpc$^{-1}$, $\Omega_{\rm m}=0.3$, and
$\Omega_\Lambda = 0.7$ is assumed throughout this Letter.

\section{Observations and data analysis\label{observations}}

We obtained deep $V$ and $R$ band images containing the \xrt\
error circle with the FORS instruments on the Kueyen and Antu 
8.2-m unit telescopes of the ESO Very Large Telescope at several 
epochs (Table~\ref{t:obslog}).
The data obtained 8 and 12 days after 
burst
%the GRB trigger 
were strongly affected by the proximity of the Moon. 
Consequently,
our deepest images were obtained during the first and last sets of
%observations (i.e., at a few days and $\sim$3 weeks after the GRB).
observations (a few days and $\sim$3 weeks after burst).

The data were bias subtracted, overscan corrected and flatfielded 
using morning skyflats. The photometric calibration was based on known 
FORS2 zero points and verified against the SDSS and photometry of
the field obtained at the Tautenburg observatory.  We estimated the
field limiting magnitudes by doing
photometry on a large ($\sim50$) number of objects. 
We used the {\tt IRAF} task {\tt phot} with an aperture of twice the
seeing disk, and obtained the $3\sigma$ limiting magnitudes given in
Table~\ref{t:obslog} from the errors on the derived magnitudes.

\begin{figure}
\epsscale{1.00}
\plotone{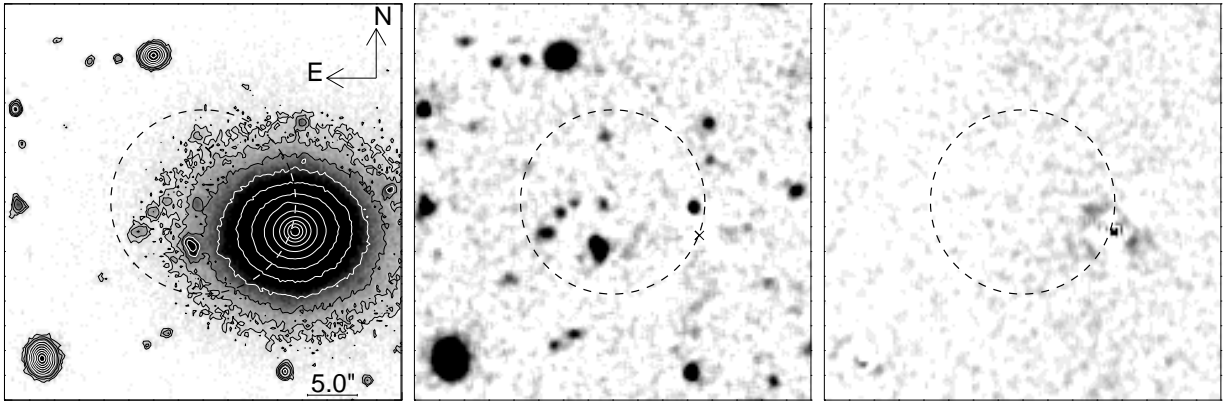}
\caption{{\it Left}: First epoch $V$ image (3.92 days after burst)
showing the putative host elliptical galaxy G1 and several faint galaxies 
in the XRT error circle.
{\it Middle}: Same, after subtraction of a fit to G1. The cross marks the
location of the center of G1. North of it is the new detected source
which may be a foreground or background source or a companion to G1.
{\it Right}: Difference between last (22.86 days after burst) and first epoch 
$V$ images.
%Plots for all sources are available
%in the electronic edition of {\it The Astrophysical Journal}.
\label{f:allimages}}
%\label{f:xrtpos}}
\end{figure}

The first image, obtained 1.85 days after the burst, revealed a large
number of very faint objects inside the \xrt\ error circle, as well as G1 
\citep[see Fig.~\ref{f:allimages} and][]{hjorth05,gehrels05,bloom05}.
The surface brightness of G1 within the
error circle is however only at a level of 20\% of the nightsky level
in both the $V$ and the $R$ bands, with the exception of the central 
1\arcsec\ of G1, corresponding to less than 1~\% of the error circle.
The pixel-to-pixel photon noise is therefore only marginally affected by G1.

To search for variable objects we subtracted the early images
from the late images in each band.  The images were
aligned, the sky background subtracted, and the images scaled
to the same brightness level. We also convolved the image with the
best seeing with a spatially variable kernel to match the inferior
seeing of the other image, according to the method outlined in
\citet{alard98}. This provided very clean subtractions, except for
near the center of the galaxy (Fig.~\ref{f:allimages}).
No variable sources were detected.

The elliptical galaxy G1 has a simple morphology. It is therefore
possible to fit and subtract a smooth model. This allows
detection of superimposed point sources with a pixel-to-pixel noise
which is reduced by a factor $\sqrt{2}$, formally corresponding to a 
0.38 mag deeper limit. A smooth fit of G1 was established by dividing 
the area around the galaxy into annuli of increasing widths, centered on
the nucleus of the galaxy. A robust fitting technique was used to fit
a harmonic series to pixel values within each of the annuli
%\citep{thomsen89,sodemann94}. The full
\citep{sodemann94}. The full
model was obtained by cubic spline interpolation between the harmonic
coefficients in the radial direction. Finally the smooth model was
subtracted from the galaxy image (Fig.~\ref{f:allimages}).
In the subtracted image we only detect one new faint ($V\sim26$) object
inside the XRT error circle, 0\farcs78 east and 2\farcs56 north of the 
G1 galaxy center. This object does not appear to be variable.
No other object brighter than the already known sources \citep{bloom05} 
is present inside the XRT error circle.

To determine the detection limit in the images, a point-spread function 
was constructed from stars in the field. By superimposing artificial 
stars on G1, we find that stars of $V\simeq26.5$
and $R\simeq25.1$ are recovered in the difference images with 95\%
confidence, except within the central 1\arcsec\ of the galaxy core.
In the very nucleus of G1, the galaxy subtraction is poor and we can 
only detect a source of $V\sim24$.  The corresponding limits for the 
galaxy subtracted images are 0.3--0.4 mag deeper.

\section{Discussion\label{discussion}}

\begin{figure}
%%\plottwo{f2.eps}{f2_color.eps}
\epsscale{.80}
\plotone{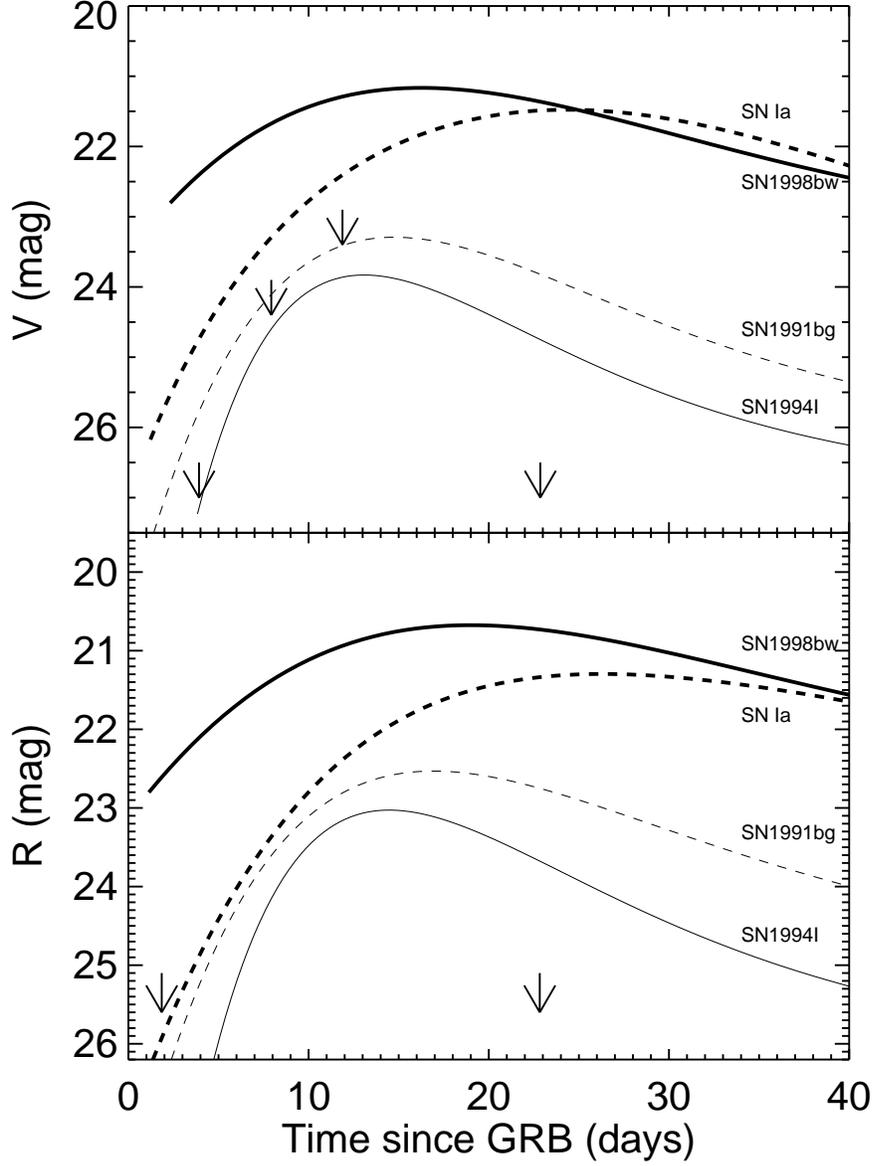}
\caption{
The upper limits ({\it arrows}) on variable sources (in the
difference images) inside the 
\grb\ XRT error circle at the epochs given in Table~1 
compared to the lightcurves of different
SNe redshifted to $z=0.225$. {\it Solid curves} indicate 
Type Ic SNe, {\it dashed curves} Type Ia SNe. 
{\it Thick solid curve}: The
hypernova SN 1998bw accompanying the long GRB 980425. {\it Thin solid curve}:
The faint Ic supernova SN 1994I. {\it Thick dashed curve}: A typical Type Ia
SN (stretch = 1). {\it Thin dashed curve}: A faint Type Ia SN
similar to SN 1991bg. A Galactic extinction of $E(B-V) = 0.019$~mag 
\citep{schlegel98} towards \grb\ has been assumed.
%See the electronic edition of the Journal for a color version 
%of this figure.
\label{f:sne}}
\end{figure}

We plot the limits derived in \S~\ref{observations} in
Fig.~\ref{f:sne} along with a number of SN lightcurves as they
would appear at $z=0.225$. The Type Ia templates are from Nugent's
compilation\footnote{http://supernova.lbl.gov/$\sim$nugent/nugent\_templates.html}
and include (i) a template of a normal Type Ia SN
\citep{nugent02} and (ii) a template based on the
%faint 
very sub-luminous Type Ia supernovae SN~1991bg and SN~1999by. The Type
Ic SNe plotted are (iii) the very energetic Type Ic SN~1998bw
associated with the long GRB 980425 \citep{galama98} and (iv) the
faint, fast-rise Type Ic SN~1994I \citep{richmond96}, which was not
associated with a GRB but provides a good fit to the lightcurve bump
in XRF 030723 \citep{fynbo04}.  Figure 2 clearly demonstrates that
even the faintest of these SNe
%(SN1994I and SN1991bg) 
would have been detected at the time of our observation at a level
$\sim1.8$~mag brighter than our limit ($>5.2$~mag fainter than a SN 
like SN~1998bw).

Type II SNe come in various flavors, the faintest of which are
Type IIP. Our limit of $V=26.5$ translates into $M_B > -13.3$ 
at $z=0.225$. All the SN peak magnitudes included in
\citet{richardson02} are brighter than this, including the
faintest Type IIP SNe. 
%However, a potential caveat here is that a
%slowly rising/decaying Type IIP SN might have had comparable
%brightness at the first and last epochs and thus would have been
%subtracted out. 

From the above we conclude that if \grb\ were associated with a normal
bright SN, its host galaxy must either be at a high redshift ($z \ga
1.2$), consistent with the constraints on the redshifts of the faint
galaxies in the \xrt\ error circle \citep[see][]{bloom05} or, if it
indeed is at $z=0.225$, its SN light must have been extinguished by
dust along the line of sight. The latter option appears unlikely as G1
is an elliptical galaxy with very little star formation
\citep{bloom05,gehrels05} and the likely background sources do not
appear strongly reddened.  We can therefore conclude that there was
no SN of known type and characteristics associated with \grb\
if it occurred in G1.  
There remains a (small) probability that \grb\ was at a much higher 
redshift.
%, possibly gravitationally lensed by G1, since the critical 
%curve (Einstein radius $r_E\approx 1\farcs5$) overlaps the \xrt\ 
%error circle \citep{pedersen05}. 
However, in the following we 
proceed under the assumption that G1 is the host galaxy.

The absence of a SN rules out models predicting a normal SN Ia
associated with short GRBs \citep{dar04,dado05}. Likewise, our
observations disfavour a \grb\ progenitor similar to long GRBs,
i.e., a collapsar origin.  Observations of long GRBs at $z<0.7$ are
consistent with all having SN bumps \citep{zeh04} and all GRBs
at $z < 0.4$ have had 
SN features [GRB\,980425,
$z = 0.0085$, \citet{galama98}; GRB\,031203, $z = 0.1055$,
\citet{malesani04}; GRB\,030329, $z=0.1685$, \citet{hjorth03};
GRB\,011121, $z=0.362$, \citet{garnavich03}].
%; GRB\,990712, $z=0.434$, \citet{bjornsson01}].  
The situation is less clear regarding X-ray
flashes (XRFs); a bright SN was associated with XRF\,020903
($z=0.251$), but no SN (and no optical afterglow) was detected in 
XRF\,040701 (with a probable $z=0.215$) down to a limit at least 
3 mag fainter than SN\,1998bw \citep{soderberg05}.

We now proceed to use our derived limits to constrain the energetic
properties of the outflow from \grb.  \citet{bloom05} find that
both the isotropic equivalent energy output in $\gamma$-rays,
$E_{\rm\gamma,iso}$, and the afterglow X-ray luminosity, $L_X$, of
\grb\ are significantly smaller than those of long GRBs. 
%This is true for any reasonable redshift, and more dramatically so for 
%the redshift of G1 ($z=0.225$). 
The most straightforward
conclusion is that, while \grb\ was a highly relativistic event 
\citep{gehrels05}, it was intrinsically less energetic compared to long 
GRBs, with relatively little energy [$\lesssim
10^{49}(\Omega/4\pi)\;$erg] in highly relativistic ejecta with an
initial Lorentz factor $\Gamma_0\gtrsim 100$ (from
$E_{\rm\gamma,iso}$) and with not much more energy in material with
$\Gamma_0\gtrsim 3(E_{51}/n_0)^{1/8}$, 
%\citep[from the {\it Chandra} upper
%limit at $t\approx 2.5\;$days;][]{patel05), 
where $E_{\rm
k,iso}=10^{51}E_{51}\;$erg is the isotropic equivalent energy in the
afterglow shock and $n=n_0\;{\rm cm^{-3}}$ is the external
density.\footnote{A higher $E_{\rm k,iso}$ is possible for a very low
external density. For $n\sim 10^{-6}\;{\rm cm^{-3}}$, $E_{\rm
k,iso}\sim 10^{51}\;$erg for $z=0.225$ and $\sim 10^{52-53}\;$erg
for $z\approx 3$ \citep{bloom05,LR-RG05}. This would, however,
require $E_{\rm\gamma,iso}\ll E_{\rm k,iso}$, i.e., a very
inefficient prompt emission (compared to $E_{\rm\gamma,iso}\gtrsim
E_{\rm k,iso}$ for long GRBs), and would not naturally reproduce the
fact that $L_X/E_{\rm\gamma,iso}$ for \grb\ is similar to that for
long GRBs.}  Moreover, the total observed energy from the burst was
much smaller than the available energy in a NS-NS or NS-BH merger, or
in most other progenitor models, suggesting that more energy was
released in slower ejecta. The amount of energy in material above a
certain initial four-velocity, $E(>\Gamma_0\beta_0)$, is very
uncertain theoretically, but may be constrained by our late time upper
limit on the optical emission.

There are two ways to produce the most readily detectable emission
from the outflow associated with \grb. It can either originate in the
shock created by the outflow as it drives into the ambient medium,
similar to both a long GRB afterglow for relativistic ejecta and to a
SN remnant for Newtonian ejecta. Or, bright transient emission,
dubbed a `mini SN' \citep{LP98,rosswog02}, is produced by
radioactive elements that are synthesized during the rapid
decompression of very dense and neutron rich material that is ejected
during a NS-NS or a NS-BH merger \citep[see, e.g.,][]{Rosswog99}. Our
upper limit at $t\approx22.8\;$days constrains mainly the former mechanism,
in particular the amount of energy in ejecta with $\Gamma_0\gtrsim
1.3(E_{51}/n_0)^{1/8}$, and suggests that the total energy in a
relativistic outflow is significantly smaller in \grb\ than that in
typical long GRBs.\footnote{One possible caveat is the dependence of
  the afterglow brightness on the density of the burst environment
  \citep[see][]{bloom05}; since the possible counterpart location on
  G1 spans a large range of densities, we have not folded this
  dependence in our conclusions.} Our upper limit at $t=1.85\;$days
provides more stringent constraints on the latter mechanism, in which
the emission is expected to peak around the optical-UV range within a
day or so (up to a few days) with a semi-thermal spectrum
\citep{LP98}. The `mini SN' emission is mainly concentrated in
a very narrow energy range (i.e., the optical) during (and near) the
peak; therefore, the X-ray emission could have been easily missed by
the {\it Chandra} observation of \grb\ at $t\approx2.5\;$days.

Using the simplified model of \citet{LP98}, the optical flux from a
`mini-SN' associated with \grb\ should have been a factor of
$\sim10^3(f/10^{-3})(M/0.01M_\odot)^{1/2}(3v/c)^{1/2}$ higher than our
upper limit at $t=1.85\;$days, where $M$ and $v$ are the mass and
velocity of the ejected material, and $f$ is the fraction of its rest
energy that goes into radioactive decay of the elements with a decay 
timescale comparable to the time it takes the expanding ejecta to 
become optically thin, $t_\tau$. For a kinetic energy of
$10^{51}E_{51}\;$erg, where $E_{51}=(M/0.01M_\odot)(3v/c)^2\sim 1$,
varying $M$ and $v$ within a reasonable range ($0.003\lesssim
M/M_\odot\lesssim 1$ and $0.03\lesssim v/c\lesssim 0.5$) would not
change the optical luminosity by more than one order of magnitude.  A
larger uncertainty is the value of $f$, which reflects the amount of
radioactive material synthesized in the accompanying NS-NS wind. From
the above simple arguments we derive an approximate upper limit of
$f\lesssim 10^{-5}$.  
We note here that the most efficient conversion of nuclear energy to the 
observable luminosity is provided by the elements with a decay timescale 
comparable to $t_\tau$. In reality, there is likely to be a large number of 
nuclides with a very broad range of decay timescales. Our limits thus place 
interesting constraints on the abundances and the lifetimes of the 
radioactive nuclides that form in the rapid decompression of nuclear-density 
matter -- they should be either very short or very long when compared to 
$t_\tau$ so that radioactivity is inefficient in generating a large 
luminosity.

The above arguments suggest that either the intrinsic energy in the
outflow from \grb\ was $\ll 10^{51}\;$erg, or alternatively, and
arguably more likely, that it was close to the canonical value of
$\sim 10^{51}\;$erg but most of this energy was in sub-relativistic
ejecta with a very small radioactive component during the optically thick 
expansion phase.\footnote{Such sub-relativistic velocities could be the result
of a significant entrainment of baryons into the outflow.}
The latter is very different from
long/soft GRBs which typically have $\sim 10^{51}\;$erg in highly
relativistic ejecta with $\Gamma_0\gtrsim 100$. 
%We note, however, that
%we need to obtain more short GRB afterglows to establish whether the
%energetics of \grb\ is representative of the bulk of the short/hard GRB class.
More short GRB afterglows are needed to establish whether the
energetics of \grb\ is representative of the bulk of the short/hard GRB class.

%UD \citep{dl98,rm00}

Finally, our observations may place constraints on other possible models 
for short GRBs
\citep[e.g.,][]{rr04,Tan01,rrrd03,Shibata03,Blackman98,Rosswog04,Davies05}.
The strict optical upper limits derived in this Letter, argue that these 
scenarios are only feasible if the transport of the energy is in the form 
of subrelativistic ejecta with little or almost no radioactivity, or in any 
other form of delayed energy input such as provided by a pulsar or by 
later mass ejection by a central source.

%Finally, our observations may place constraints on other possible models 
%for short GRBs.  For instance, the central object may not become 
%dormant after the GRB itself
%\citep[e.g.,][]{rr04}. It could be
%that the accretion-induced collapse of a white dwarf \citep{Tan01}, or
%(for some equations of state) 
%the merger of two NS, could
%give rise to a rapidly-spinning pulsar \citep{rrrd03}, temporarily
%stabilized by rapid rotation. The afterglow could then, at least in
%part, be due to a pulsar's continuing power output
%\citep{dl98,rm00}. It could also be that mergers of unequal mass
%NS \citep{Shibata03,Blackman98}, or NS with
%other compact companions \citep{Rosswog04,Davies05}, lead to the
%delayed formation of a BH. Such events might also lead to
%repeating episodes of accretion and orbit separation, or to the
%eventual explosion of a NS which has dropped below the
%critical mass, all of which would provide a longer time scale, and
%episodic energy output. The strict optical upper limits derived in
%this Letter, argue that these scenarios are only feasible if the
%transport of the energy is in the form of subrelativistic ejecta with
%little 
%or almost no
%radioactivity, or in any other form of delayed energy input. 
%such as provided by a pulsar or by later mass ejection by a central source.

\acknowledgments

We thank Elena Pian and Alberto Castro-Tirado for comments and 
the ESO Paranal Science Operations staff, in particular Chris
Lidman and Andreas O. Jaunsen for efficiently conducting the
observations reported here. 
The Dark Cosmology Centre is supported by the DNRF.
The authors acknowledge benefits from
collaboration within the EC FP5 Research Training Network ``Gamma-Ray
Bursts -- An Enigma and a Tool''.
%, contract number HPRN-CT-2002-00294. 
%The research of J.~G. is supported by the US Department of Energy under 
%contract number DE-AC03-76SF00515.


%% copy editing.  Individual instruments can be provided in parentheses,
%% after the keyword, but they will not be verified.

Facilities: \facility{VLT(FORS1,FORS2)}


{}

\clearpage

%% the file, uncomment the \plotone, \plottwo, and
%% If you need a layout that cannot be achieved with \plotone or
%% \plottwo, you can invoke the graphicx package directly with the
%% \includegraphics command or use \plotfiddle. For more information,
%% please see the tutorial on "Using Electronic Art with AASTeX" in the
%% documentation section at the AASTeX Web site,
%% http://www.journals.uchicago.edu/AAS/AASTeX.
%%
%% The examples below also include sample markup for submission of
%% supplemental electronic materials. As always, be sure to check
%% the instructions to authors for the journal you are submitting to
%% for specific submissions guidelines as they vary from

%% This example uses \plotone to include an EPS file scaled to
%% 80% of its natural size with \epsscale. Its caption
%% has been written to indicate that additional figure parts will be
%% available in the electronic journal.

\clearpage

%% Here we use \plottwo to present two versions of the same figure,
%% one in black and white for print the other in RGB color
%% for online presentation. Note that the caption indicates
%% that a color version of the figure will be available online.
%%

%\begin{figure}[t]
%\setlength{\unitlength}{1mm}
%\begin{picture}(120,120)(0,0)
%\put (0,59) {\includegraphics[width=58mm,bb=36 125 577 668,clip]{Vepoch1.ps}}
%\put (59,59){\includegraphics[width=58mm,bb=36 125 577 668,clip]{Repoch4.ps}}
%\put (0,0) {\includegraphics[width=58mm,bb=36 125 577 668,clip]{V1bjarne.ps}}
%\put (59,0){\includegraphics[width=58mm,bb=36 125 577 668,clip]{Vdiffjens.ps}}
%\end{picture}
%\caption{
%{\it Upper left:}
%V band image obtained 4 days past the burst. The field of view is 
%$40\arcsec \times 40\arcsec$ and North is up and East to the left for all the 
%sub-images. A large fraction of the elliptical galaxy G1 is 
%inside the \xrt\ error circle.
%{\it Upper right:}
%R band image obtained 23 days past burst.
%{\it Lower left:}
%This image shows the V band image above, after subtraction of the elliptical galaxy.
%{\it Lower right:}
%This image shows the subtraction in the V band between the first and last
%of our epochs. No variable sources are detected.
%\label{f:allimages}}
%\end{figure}

%% This figure uses \includegraphics to scale and rotate the still frame

%\begin{figure}
%%%\includegraphics[angle=90,scale=.50]{f3.eps}
%\caption{Animation still frame taken from \citet{kim03}.
%This figure is also available as an mpeg
%animation in the electronic edition of the
%{\it Astrophysical Journal}.}
%\end{figure}

%% If you are not including electonic art with your submission, you may
%% mark up your captions using the \figcaption command. See the
%% No more than seven \figcaption commands are allowed per page,
%% so if you have more than seven captions, insert a \clearpage

%% Tables should be submitted one per page, so put a \clearpage before

%% Two options are available to the author for producing tables:  the
%% deluxetable environment provided by the AASTeX package or the LaTeX
%% table environment.  Use of deluxetable is preferred.
%%

%% Three table samples follow, two marked up in the deluxetable environment,
%% one marked up as a LaTeX table.

%% In this first example, note that the \tabletypesize{}
%% command has been used to reduce the font size of the table.
%% We also use the \rotate command to rotate the table to
%% landscape orientation since it is very wide even at the
%% Note also that the \label command needs to be placed

%% This table also includes a table comment indicating that the full
%% version will be available in machine-readable format in the electronic
\clearpage

\begin{deluxetable}{lccccc}
%\tabletypesize{\scriptsize}
%\rotate
\tablecaption{Log of observations\label{t:obslog}}
\tablewidth{0pt}
\tablehead{
\colhead{Date} & \colhead{Phase} & \colhead{Band} & \colhead{Exp.\ time} & \colhead{Seeing} & \colhead{Lim.\ mag}\\
\colhead{(UT)} & \colhead{(Days past GRB)} & \colhead{} & \colhead{(s)} & \colhead{(arcsec)} & \colhead{(mag)}
}
\startdata
050511.02 & 1.85          & $R$  & 2700    & 0.9   & 26.6    \\
050513.09 & 3.92          & $V$  & 2700    & 0.9   & 27.5    \\
050517.10 & 7.93          & $V$  & 1800    & 0.7   & 25.0    \\
050522.05 & 11.88         & $V$  & 1800    & 1.0   & 24.2    \\
050601.00 & 22.83         & $R$  & 2700    & 0.9   & 26.7    \\
050601.03 & 22.86         & $V$  & 2700    & 1.0   & 27.5    \\
\enddata

%% Text for table notes should follow after the \enddata but before
%% the \end{deluxetable}. Make sure there is at least one \tablenotemark
%% in the table for each \tablenotetext.

\tablecomments{
The quoted 3$\sigma$ limiting magnitudes are measured in 
the field in a 2$\times$FWHM aperture. 
The limiting magnitudes become progressively
smaller towards the center of G1.
%Table \ref{t:obslog} is published in its entirety in the
%electronic edition of the {\it Astrophysical Journal}. A portion is shown
%here for guidance regarding its form and content.
}

%\tablenotetext{a}{Sample footnote for table~\ref{tbl-1} that was generated
%with the deluxetable environment}
%\tablenotetext{b}{Another sample footnote for table~\ref{tbl-1}}

\end{deluxetable}


\begin{thebibliography}{}

\bibitem[Alard \& Lupton(1998)]{alard98} 
Alard, C., \& Lupton, R.~H.  1998, \apj, 503, 325 

%\bibitem[Bj\"ornsson et al.(2001)]{bjornsson01}
%Bj\"ornsson, G., Hjorth, J., Jakobsson, P., Christensen, L., \& Holland, S. T.
%2001, \apjl, 552, L121

\bibitem[Blackman \& Yi(1998)]{Blackman98} Blackman, E.~G., \& Yi, 
I.\ 1998, \apjl, 498, L31 

%\bibitem[Bloom et al.(1998)]{bloom98} 
%Bloom, J.~S., Djorgovski, S.~G., Kulkarni, S.~R., \& Frail, D.~A.  1998, 
%\apjl, 507, L25 

\bibitem[Bloom et al.(1999)]{bloom99} 
Bloom, J.~S., et al.\  1999, \nat, 401, 453 

\bibitem[Bloom et al.(2005)]{bloom05} 
Bloom, J.~S., et al.\  2005, \apj, submitted (astro-ph/0505480)

%\bibitem[Bloom, Kulkarni \& Djorgovski(2002)]{bloom02} 
%Bloom, J. S., Kulkarni, S. R., \& Djorgovski, S. G.  2000, \aj, 123, 1111

\bibitem[Castro-Tirado et al.(2005)]{alberto05} 
Castro-Tirado, A. J., et al.\  2005, \aap, 439, L15
%in press (astro-ph/0506662)

\bibitem[Cenko et al.(2005)]{cenko05} 
Cenko, S. B., Soifer, B. T., Bian, C., Desai, V., Kulkarni, S. R.,
Berger, E., Dey, A., \& Jannuzi, B. T.  2005, GCNC 3391

%\bibitem[Christensen, Hjorth \& Gorosabel(2004)]{christensen04} 
%Christensen, L., Hjorth, J., \& Gorosabel, J.   2004, \aap, 425, 913

%\bibitem[Dai \& Lu(1998)]{dl98} Dai, Z.~G., \& Lu, T. 1998,
%Phys.\ Rev.\ Lett., 81, 4301

\bibitem[Dado, Dar \& De R\'ujula(2005)]{dado05} 
Dado, S., Dar, A., \& De R\'ujula, A.  2005, GCNC 3424

\bibitem[Dar \& De R\'ujula(2004)]{dar04} 
Dar, A., \& De R\'ujula, A.  2004, Phys.~Rep., 405, 203

\bibitem[Davies, Levan \& King(2005)]{Davies05} 
Davies, M. B., Levan, A. J., \& King, A. R.  2005, \mnras, 356, 54

%\bibitem[Engelbracht \& Eisenstein(2005)]{Engelbracht05} 
%Engelbracht, C.W. \& Eisenstein, D.J. 2005, GCNC 3420

\bibitem[Fan et al.(2005)]{fan05} 
Fan, Y. Z., Zhang, B., Kobayashi, S., \& M\'esz\'aros, P.  2005, \apj, 628, 867
%in press (astro-ph/0410060)

%\bibitem[Fruchter et al.(1999)]{fruchter99} 
%Fruchter, A. S., et al.\  1999, \apjl, 519, 13

\bibitem[Fynbo et al.(2004)]{fynbo04} 
Fynbo, J.~P.~U., et al.\  2004, \apj, 609, 962 

\bibitem[Gal et al.(2003)]{gal03} 
Gal, R. R., de Carvalho, R. R., Lopes, P. A. A., Djorgovski, S. G.,
Brunner, R. J., Mahabal, A., \& Odewahn, S. C.  2003, \aj, 125, 2064 

\bibitem[Galama et al.(1998)]{galama98} 
Galama, T.~J., et al.\ 1998, \nat, 395, 670 

\bibitem[Garnavich et al.(2003)]{garnavich03} 
Garnavich, P. M., et al.\  2003, \apj, 582, 924

\bibitem[Gehrels et al.(2004)]{gehrels04}
Gehrels, N., et al.\  2004, \apj, 611, 1005 

\bibitem[Gehrels et al.(2005)]{gehrels05}
Gehrels, N., et al.\  2005, Nature, in press (astro-ph/0505630)

\bibitem[Germany et al.(2000)]{germany00}
Germany, L. M., Reiss, D. J., Sadler, E. M., Schmidt, B. P., \& Stubbs, C. W.
2000, \apj, 533, 320

\bibitem[Ghirlanda, Ghisellini \& Celotti(2004)]{ghirlanda04}
Ghirlanda, G., Ghisellini, G., \& Celotti, A.  2004, \aap, 422, L55

\bibitem[Gorosabel et al.(2002)]{gorosabel02}
Gorosabel, J., et al.\  2002, \aap, 383, 112

\bibitem[Hjorth et al.(2003)]{hjorth03} 
Hjorth, J., et al.\  2003, \nat, 423, 847 

\bibitem[Hjorth et al.(2005)]{hjorth05} 
Hjorth, J., et al.\  2005, GCNC 3410

\bibitem[Hurley et al.(2002)]{hurley02} 
Hurley, K., et al.\  2002, \apj, 567, 447

\bibitem[Kehoe et al.(2001)]{kehoe01} 
Kehoe, R., et al.\  2001, \apj, 554, L159

\bibitem[Klotz, Bo\"er \& Atteia(2003)]{klotz03} 
Klotz, A., Bo\"er, M., \& Atteia, J.-L.  2003, \aap, 815, 2003

\bibitem[Kouveliotou et al.(1993)]{kouveliotou93} 
Kouveliotou, C., Meegan, C. A., Fishman, G. J., Bhat, N. P., Briggs, M. S.,
Koshut, T. M., Paciesas, W. S., \& Pendleton, G. N.  1993, \apjl, 413, 110 

\bibitem[Lee, Ramirez-Ruiz \& Granot(2005)]{LR-RG05} 
Lee, W.~H., Ramirez-Ruiz, E., \& Granot, J. 2005, \apj, submitted
  (astro-ph/0506104)

%\bibitem[Le Floc'h et al.(2003)]{lefloch03} 
%Le Floc'h, E., et al.\ 2003, \aap, 400, 499

%\bibitem[Levan et al.(2005)]{levan05}
%Levan, A., et al.\  2005, \apj, 622, 977

\bibitem[Li \& Paczy\'nski(1998)]{LP98}
Li, L.~X., \& Paczy\'nski, B. 1998, ApJ, 507, L59

\bibitem[MacFadyen \& Woosley(1999)]{macfadyen99} 
MacFadyen, A. I., \& Woosley, S. E.  1999, \apj, 524, 262

\bibitem[Malesani et al.(2004)]{malesani04}
Malesani, D., et al.\  2004, \apjl, 609, L5

%\bibitem[Nakar et al.(2005)]{nakar05} 
%Nakar, E., Gal-Yam, A., Piran, T., \& Fox, D. B.  2005, astro-ph/0502148

\bibitem[Nugent, Kim \& Perlmutter(2002)]{nugent02} 
Nugent, P., Kim, A., \& Perlmutter, S.  2002, \pasp, 114, 803

%\bibitem[Patel et al.(2005)]{patel05} 
%Patel, S., Kouveliotou, C., Burrows, D.~N., Grupe, D., 
%Gehrels, N., M\'esz\'aros, P., Zhang, B., \& Wijers, R.  2005, GCNC 3419

%\bibitem[Pedersen et al.(2005)]{pedersen05} 
%Pedersen, K., et al.\ 2005, in preparation

%\bibitem[Prochaska et al.(2004)]{prochaska04} 
%Prochaska, J.~X., et al.\  2004, \apj, 611, 200 

\bibitem[Richardson et al.(2002)]{richardson02} 
Richardson, D., Branch, D., Casebeer, D., Millard, J., Thomas, R. C.,
\& Baron, E.  2002, \aj, 123, 745

\bibitem[Richmond et al.(1996)]{richmond96} 
Richmond, M. W., et al.\  1996, \aj, 111, 327

\bibitem[Ramirez-Ruiz(2004)]{rr04} Ramirez-Ruiz, E. 2004, \mnras,
349, L38

%\bibitem[Rees \& M{\' e}sz{\' a}ros(2000)]{rm00} Rees, M.~J., \& M{\'
%e}sz{\' a}ros, P. 2000, ApJ, 545, L73

\bibitem[Rosswog et al.(1999)]{Rosswog99} Rosswog, S., Liebend{\" 
o}rfer, M., Thielemann, F.-K., Davies, M.~B., Benz, W., \& Piran, T.\ 1999, 
\aap, 341, 499 
 
\bibitem[Rosswog \& Ramirez-Ruiz(2002)]{rosswog02} Rosswog, S., \&
Ramirez-Ruiz, E.  2002, \mnras, 336, L7

\bibitem[Rosswog, Ramirez-Ruiz \& Davies(2003)]{rrrd03} Rosswog, S.,
Ramirez-Ruiz, E., \& Davies, M.~B.  2003, \mnras, 345, 1215

\bibitem[Rosswog et al.(2004)]{Rosswog04} Rosswog, S., Speith, R., \&
Wynn, G.~A.\ 2004, \mnras, 351, 1121

\bibitem[Schlegel et al.(1998)]{schlegel98} Schlegel, D.~J.,
Finkbeiner, D.~P., \& Davis, M.  1998, \apj, 500, 525

\bibitem[Shibata \& Sekiguchi(2003)]{Shibata03} 
Shibata, M., \& Sekiguchi, Y.\ 2003, \prd, 68, 104020 

\bibitem[Sodemann \& Thomsen(1994)]{sodemann94} 
Sodemann, M., \& Thomsen, B.  1994, \aap, 292, 425

\bibitem[Soderberg et al.(2005)]{soderberg05}
Soderberg, A. M., et al.\  2005, \apj, 627, 877 % (astro-ph/0502553)

\bibitem[Stanek et al.(2003)]{stanek03} 
Stanek, K.~Z., et al.\  2003, \apjl, 591, L17

\bibitem[Tan, Matzner \& McKee(2001)]{Tan01} 
Tan, J.~C., Matzner, C.~D., \& McKee, C.~F.\ 2001, \apj, 551, 946 

%\bibitem[Thomsen \& Baum(1989)]{thomsen89} 
%Thomsen, B. \& Baum, W. A.  1989, \apj, 347, 214

%\bibitem[Thomsen et al.(2004)]{thomsen04} 
%Thomsen, B., et al.\  2004, \aap, 419, L21

%\bibitem[van den Bergh, Li \& Filippenko(2003)]{vandenbergh03} 
%van den Bergh, S., Li, W., \& Filippenko, A. V.  2003, \pasp, 115, 1280

\bibitem[Yamazaki, Ioka \& Nakamura(2004)]{yamazaki04} 
Yamazaki, R., Ioka, K., \& Nakamura, T.  2004, \apjl, 607, L103

\bibitem[Zeh, Klose \& Hartmann(2004)]{zeh04} 
Zeh, A., Klose, S., \&  Hartmann, D. H.  2004, \apj, 609, 952

%\bibitem[Zwicky \& Herzog(1963)]{zwicky63} 
%Zwicky, F., \& Herzog, E.  1963, Catalogue of Galaxies and of
%Clusters of Galaxies, Vol.~II, Pasadena: California Institute of Technology

\bibitem[Zhang, Woosley \& MacFadyen(2003)]{zhang03} 
Zhang, W., Woosley, S. E., \& MacFadyen, A. I.  2003, \apj, 586, 356

\end{thebibliography}
\end{document}